%% file: main.tex
\documentclass[sigconf,nonacm]{acmart}
\acmConference[ISSTA 2023]{ACM SIGSOFT International Symposium on Software Testing and Analysis}{17-21 July, 2023}{Seattle, USA}

\title{Automatically Reproducing Android Bug Reports Using Natural Language Processing and Reinforcement Learning}

\input{packages}
\input{macros}
\title{Automatically Reproducing Android Bug Reports Using Natural Language Processing and Reinforcement Learning}
\author{Zhaoxu Zhang}
\affiliation{%
  \institution{University of Southern California}
  \country{USA}}
\email{zhaoxuzh@usc.edu}

\author{Robert Winn}
\affiliation{%
  \institution{University of Southern California}
  \country{USA}}
\email{rwinn@usc.edu}

\author{Yu Zhao}
\affiliation{%
  \institution{University of Central Missouri}
  \country{USA}}
\email{yzhao@ucmo.edu}

\author{Tingting Yu}
\affiliation{%
  \institution{University of Cincinnati}
  \country{USA}}
\email{yutt@ucmail.uc.edu}

\author{William G.J. Halfond}
\affiliation{%
  \institution{University of Southern California}
  \country{USA}}
\email{halfond@usc.edu}
\begin{document}

\input{content/abstract}
\maketitle
\input{content/introduction}
\input{content/motivating_example}
\input{content/approach}

\input{content/evaluation}
\input{content/related_work}

\input{content/conclusion}
\bibliographystyle{acm}
\bibliography{main, group, zotero}

\end{document}

%% file: packages.tex
\usepackage[utf8]{inputenc}
\usepackage{acronym}
\usepackage[inline]{enumitem}
\usepackage{amsmath}
\usepackage{tcolorbox}
\tcbuselibrary{skins}
\usetikzlibrary{patterns, external, shadings}
\usepackage[normalem]{ulem}
\usepackage{xspace}
\usepackage{algorithm2e}
\SetKw{Continue}{continue}
\SetKw{Break}{break}
\usepackage{subcaption}
\usepackage{graphicx}
\usepackage{multirow}
\usepackage{ctable}
\usepackage{tabularx}
\usepackage{algpseudocode}
\usepackage[online,para]{threeparttable}
\usepackage{cleveref}

%% file: macros.tex
\newacro{openie}[Open IE]{open information extraction}
\newacro{ie}[IE]{information extraction}
\newacro{rl}[RL]{reinforcement learning}
\newacro{s2r}[S2R]{step to reproduce}
\acrodefplural{s2r}{steps to reproduce}
\newacro{nlp}[NLP]{natural language process}
\newacro{nl}[NL]{natural language}
\newacro{ui}[UI]{user interface}
\newacro{mdp}[MDP]{markov decision process}
\newacro{dfs}[DFS]{depth-first search}
\newacro{fp}[FP]{false positive}
\newacro{fps}[FPs]{false positives}
\newacro{fn}[FN]{false negative}
\newacro{fns}[FNs]{false negatives}
\newacro{aut}[AUT]{app under test}
\newacro{doet}[DOET]{dynamic ordered event tree}
\newacro{dpt}[DPT]{dependency parsing tree}
\newacro{cpt}[CPT]{constituency parsing tree}
\newacro{np}[NP]{noun phrase}
\newacro{pos}[POS]{part of speech}
\newacro{pp}[PP]{prepositional phrase}
\newacro{dfs}[DFS]{depth first search}
\newacro{dt}[DT]{discourse tree}
\newacro{vh}[VH]{view hierarchy}
\newacro{pdtb}[PDTB]{Peen Discourse Treebank}
\newcommand{\tool}{\textsc{ReproBot}\xspace}

\newacro{noop}[no-op]{no operation}
\newcommand{\toolvara}{\textsc{RB\textsubscript{a}}\xspace}
\newcommand{\toolvarb}{\textsc{RB\textsubscript{b}}\xspace}
\newcommand{\recdroid}{\textsc{ReCDroid}\xspace}
\newcommand{\yakusu}{\textsc{Yakusu}\xspace}

\newcommand{\euler}{\textsc{Euler}\xspace}
\newcommand{\maca}{\textsc{Maca}\xspace}
\newcommand{\fusion}{\textsc{Fusion}\xspace}
\newcommand{\ebug}{\textsc{EBUG}\xspace}
\newcommand{\burt}{\textsc{BURT}\xspace}
\newcommand{\qlearning}{Q-learning\xspace}
\newcommand{\qvalue}{Q-value\xspace}

\newcommand{\ie}{i.e.,\xspace}
\newcommand{\eg}{e.g.,\xspace}
\newcommand{\etal}{et al.\xspace}
\newcommand{\etc}{etc.\xspace}

\newcommand{\simthred}{$d$\xspace}
\newcommand{\defaultsimscore}{$r_d$\xspace}
\newcommand{\failurepenalty}{$r_f$\xspace}
\newcommand{\explorationpenalty}{$r_e$\xspace}

\def \keepnotes {}
\ifthenelse{\isundefined{\keepnotes}}{
\newcommand{\gj}[1]{}
\newcommand{\zx}[1]{}
\newcommand{\commentty}[1]{}
\newcommand{\YZ}[1]{}
\newcommand{\rw}[1]{}
}{
\newcommand{\commentty}[1]{{\color{purple} \sf (TY: #1)}}
\newcommand{\YZ}[1]{{\color{green} \sf (YZ: #1)}}
\newcommand{\gj}[1]{\textcolor{red}{\sf (GJ: #1)}}
\newcommand{\zx}[1]{\textcolor{blue}{\sf (ZZ: #1)}}
\newcommand{\rw}[1]{\textcolor{cyan}{\sf (RW: #1)}}
}

%% file: content/abstract.tex
\begin{abstract}

As part of the process of resolving issues submitted by users via bug reports, Android developers attempt to reproduce and observe the failures described by the bug report. Due to the low-quality of bug reports and the complexity of modern apps, the reproduction process is non-trivial and time-consuming. Therefore, automatic approaches that can help reproduce Android bug reports are in great need. However, current approaches to help developers automatically reproduce bug reports are only able to handle limited forms of natural language text and struggle to successfully reproduce failures for which the initial bug report had missing or imprecise steps. In this paper, we introduce a new fully automated Android bug report reproduction approach that addresses these limitations. Our approach accomplishes this by leveraging \acl{nlp} techniques to more holistically and accurately analyze the natural language in Android bug reports and designing new techniques, based on \acl{rl}, to guide the search for successful reproducing steps. We conducted an empirical evaluation of our approach on $77$ real world bug reports. Our approach achieved 67\% precision and 77\% recall in accurately extracting reproduction steps from bug reports, and reproduced 74\% of the bug reports, significantly outperforming state of the art techniques.

\end{abstract}

%% file: content/introduction.tex
\section{Introduction}
\label{sec:intro}

In the hyper-competitive world of app marketplaces, app developers strive to provide their users with interesting features and high-quality functionality to distinguish themselves from their competition.  An important mechanism for receiving feedback from users is bug reporting systems, such as GitHub Issues Tracker~\cite{GHIssue} and Google Code~\cite{GCIssue}.  These systems enable users to create \textit{bug reports} in which they can describe the observed failure and provide steps to reproduce the failure.  Developers can use this information to help debug their apps. However, the use of this information is complicated by the fact that the bug reports are often informally written in natural language, imprecise, and incomplete~\cite{moranAutocompletingBugReports2015,johnsonEmpiricalInvestigationReproduction2022,zimmermannWhatMakesGood2010}. This can make it challenging for developers to reproduce the reported failure, as important steps may be missing or poorly described.  Even with well-written bug reports, reproduction can still be challenging since mobile apps often have complex event-driven user interfaces that allow many similar sequences of actions, each of which may or may not lead to the reported failure. Taken together, these aspects can make bug report reproduction a time-consuming and error-prone process, which can undermine the usefulness of the bug reporting process.

The software engineering community has tried to address this problem through the development of automated bug report reproduction techniques (\eg \cite{fazziniAutomaticallyTranslatingBug2018,zhaoAutomaticallyExtractingBug2019,zhaoReCDroidAutomatedEndtoEnd2022}).  These approaches generally have two phases.  In the first phase (i.e., bug report analysis), the approaches analyze the natural language in the bug reports in order to identify the \acp{s2r}. 
Each \ac{s2r} describes an action on a \ac{ui} element in the \ac{aut}.  In the second phase (i.e., App exploration), the reproduction approaches attempt to execute each \ac{s2r} on the \ac{aut}. 
Both phases represent significant challenges that make it difficult to fully automate this process.  In the first phase, the natural language is generally unstructured, written by users without a technical background, and has similar concepts described in a multitude of ways~\cite{liuAutomatedClassificationActions2020, fazziniAutomaticallyTranslatingBug2018}.  Even if the first phase could be done perfectly, many bug reports have missing steps~\cite{johnsonEmpiricalInvestigationReproduction2022, wendlandAndror2DatasetManuallyReproduced2021}.  This complicates the second phase, since the approaches must either find some ways to identify plausible missing steps or dead end in their reproduction efforts when no \ac{ui} element matches the next \ac{s2r}.  This happens, for example, when a missing \ac{s2r} specifies a step that causes a \ac{ui} element to appear that is then used by the subsequent \acp{s2r} in the bug report. 

Two state of the art approaches, \yakusu~\cite{fazziniAutomaticallyTranslatingBug2018} and \recdroid~\cite{zhaoReCDroidAutomaticallyReproducing2019}, define techniques for handling these challenges. To address the challenge in the first phase, both approaches extract \acp{s2r} from bug report text using manually crafted patterns and predefined word lists that map to standard actions.  For example, for the phrase "click the Home button," these approaches would identify a \textit{click} with the "Home button" as the target.  
However, these techniques are unable to handle natural language in bug reports with either previously unseen words or different sentence structures.
\maca~\cite{liuAutomatedClassificationActions2020} designs a classifier to normalize action words into a standard form, but also uses simple patterns to parse the sentence and therefore has limitations when handling previously unseen sentence structures.
To address the challenge in the second phase, both approaches employ a greedy strategy to explore the app and identify possible mappings of \acp{s2r} to UI events. 
However, the greedy strategy can lead the approaches to prioritize matching an \ac{s2r} with a UI event that is the most similar, even though choosing a lower similarity match may allow subsequent \acp{s2r} to match better with other UI events. This may occur, for example, when there are inaccurately described or missing steps. 



To address the aforementioned challenges, we designed a new bug report reproduction approach.  At a high-level, our approach follows the two-phase architecture defined by prior techniques.  However, for each phase, we developed new algorithms and designs that enable our approach to be \textit{more broadly applicable} -- able to handle a wider variety of natural language in bug reports -- and \textit{more successful} in matching \acp{s2r} extracted from the bug report to actions in an app's \ac{ui}.  
We achieved the first of these by developing a set of analyses that extract \ac{s2r} information without relying on predefined patterns.
To improve reproduction, we formulate the task of matching \acp{s2r} with an app's \ac{ui} events into a \ac{mdp}. Our approach then identifies the reproduction event sequence by leveraging \ac{rl} algorithms, specifically Q-learning~\cite{q-learning}.  
Our approach employs Q-learning to learn how to match the \ac{ui} events with the \acp{s2r} in such a way as to bridge missing steps and calculate an overall best match between \acp{s2r} and a \ac{ui} event sequence that can lead to the observed failure. 
We implemented our approach as a prototype tool and compared it against \recdroid and \yakusu.  Our results show that our \ac{nlp} techniques help our approach to handle a wider variety of bug reports and our \ac{rl} based exploration leads to a more successful reproduction phase. Together, these two techniques enable our approach to outperform \recdroid and \yakusu by a significant margin and indicate that our approach represents a significant improvement to state of the art bug report reproduction techniques.

In summary, our paper makes the following contributions:
\begin{itemize}
    \item We designed a novel NLP based analysis to analyze and extract reproduction steps from Android bug reports.
    \item We designed a new exploration strategy, 
     based on \acl{rl}, for exploring Android apps and finding the best overall match between steps and UI elements.
    \item Based on the above two contributions, we developed a novel approach to reproduce Android bug reports and implemented it as a tool.
    \item We conducted an empirical evaluation showing the performance of our approach.
    \item We will make the implementation and dataset publicly available for future research work.
\end{itemize}

%% file: content/motivating_example.tex
\section{Motivating Example}
\label{sec:motivating_example}

\input{content/pictures/example}
In this section we introduce a motivating example that we will use throughout the paper to illustrate our approach and highlight the limitations of existing approaches. 
The bug report, which is shown in \Cref{fig:bugreport}, is based on bug reports that we found in our evaluation subjects. The bug report describes a failure that occurs when the user tries to take a photo on the preference page. It reports four steps (as annotated) that need to be taken for reproduction. Although this report seems reasonably clear and is, in fact, typical of most bug reports, it can cause several problems for the state of the art automated bug report reproduction techniques (e.g., \recdroid~\cite{zhaoReCDroidAutomaticallyReproducing2019}, \yakusu~\cite{fazziniAutomaticallyTranslatingBug2018}) when they try to reproduce it.
To reproduce a bug report, these automated techniques first analyze the natural language of the bug report to determine the \acp{s2r}. 
Each \ac{s2r} is a structured form that contains entities describing the step such as \ac{ui} action and target widget.
Although the text is seemingly easy to parse (for a human), these automated techniques would run into several problems when trying to do so automatically.  Both techniques try to identify an \ac{s2r} by matching text in the bug report to a pre-defined vocabulary of words describing \ac{ui} actions.  
This would make them fail to extract the first step in our example, which uses a verbal phrase ``attempt to take'' to express the \ac{ui} action, an unusual way to describe a \ac{ui} action that is not present in either approach's vocabulary.
Second, after identifying a \ac{ui} action, both techniques rely on hand-crafted rules to match other \ac{s2r} entities.
However, none of these rules are able to identify the target ``photo''. 
Alternatively, a misidentification may also happen if the rules are too simple.  For example, \recdroid would identify the target of the fourth step to be ``CANCEL'' and ``circle''. 
These inaccurate entity identifications would create noise when the automated techniques attempt to match the \acp{s2r} in the \ac{aut}.  
Additionally, both approaches suppose the execution order of \acp{s2r} is the same as their syntactic order in the text. 
However, this could lead both of them to obtain the wrong order of the first two steps.
As indicated by the semantics of the connective word "when" between the first two steps, the second step should happen first and then be followed by the first step.

Even assuming there is no noise in the \acp{s2r}, these approaches would face additional challenges during reproduction when they try to automatically match \acp{s2r} against \ac{ui} events. First, both techniques employ a greedy approach when matching the \acp{s2r} with \ac{ui} events, meaning that they always choose the event that is the most closely matched with the \ac{s2r}. This could lead to a local maximum in the matching process, effectively trapping the exploration and preventing it from finding a better overall match. Second, both techniques assume that missing steps only occur when there is no \ac{ui} event matched with the \acp{s2r} during the exploration. However, the missing steps could occur exactly when there are events that match the \acp{s2r}. 
When the above cases happen, prior reproduction tools would either fail to reproduce the failure or reduce to an exhaustive exploration of the \ac{aut}.

These limitations of the state of the art directly motivate the design of our approach.  Given a bug report sentence, our approach first normalizes the execution order of \acp{s2r} in the sentence. 
To identify the entities for a \ac{s2r}, instead of using predefined patterns, we designed an analysis that infers them from a set of more general syntactic parts that could be identified from any natural language sentence.
These two steps together allow our approach to have a more accurate understanding of the bug report sentence and are also more widely applicable. 
To match \acp{s2r} and \ac{ui} events, we designed an exploration and matching strategy based on \ac{mdp} and \qlearning. These techniques enable our exploration approach to find a way to bridge possible gaps in the \acp{s2r} and at the same time avoid local search maxima (i.e., that would be found with a greedy approach).  In the next section, we explain our approach's design and algorithms for doing this in more detail.
%


%% file: content/pictures/example.tex
\tcbset{enhanced,colback=black!3!white, colframe=black!75!black}

\begin{figure}[h]
\centering
\includegraphics[scale=0.43]{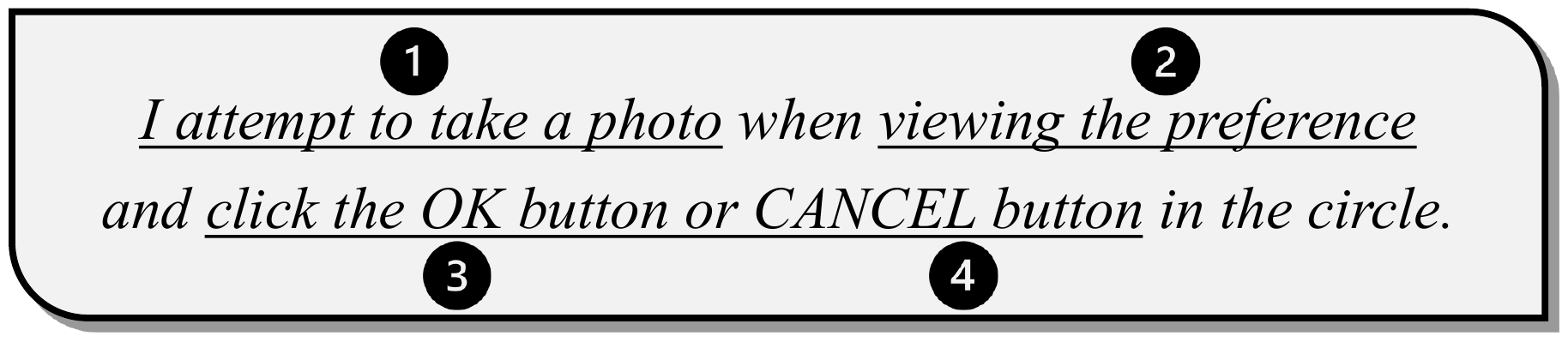}
\caption{Bug Report Example}
\label{fig:bugreport}
\end{figure}

%% file: content/approach.tex
\section{Approach}
\input{content/pictures/workflow}

The goal of our approach is to automatically reproduce app failures described in bug reports. The input of our approach is \ac{s2r} sentences - \textit{natural language sentences describing reproduction steps}, and the \ac{aut}. 
At a high-level, \Cref{fig:workflow} shows the two stages of our approach: Extraction of \ac{s2r} Entities (\ac{s2r} Extraction) and Matching \acp{s2r} to \ac{ui} Events (\ac{s2r} Matching).  
In the first stage, our approach analyzes the \ac{s2r} sentences for the purpose of extracting the \ac{s2r} entities defining each step to reproduce the bug report.
The entities we extracted include information such as the type of \ac{ui} action to be performed and the target of the action.
Our approach leverages a combination of natural language techniques to systematically analyze \ac{s2r} sentences in order to extract the entities more accurately. 
In the second stage, our approach explores the app to match the reproduction steps to \ac{ui} events in the \ac{aut}.
We formulate the task of matching \acp{s2r} entities to UI events as a \acf{mdp} and use Q-learning algorithm (\Cref{sec:approach:reproduction}) to identify the best matching relation that leads to the failure.
The combination of \ac{mdp} and \qlearning allows our approach to effectively bridge missing steps and identify a globally optimal match sequence that leads to the observed failure.




\input{content/approach_s2r_extraction}

\input{content/approach_s2r_matching}

%% file: content/pictures/workflow.tex
\begin{figure}[t]
\centering
\includegraphics[scale=0.51]{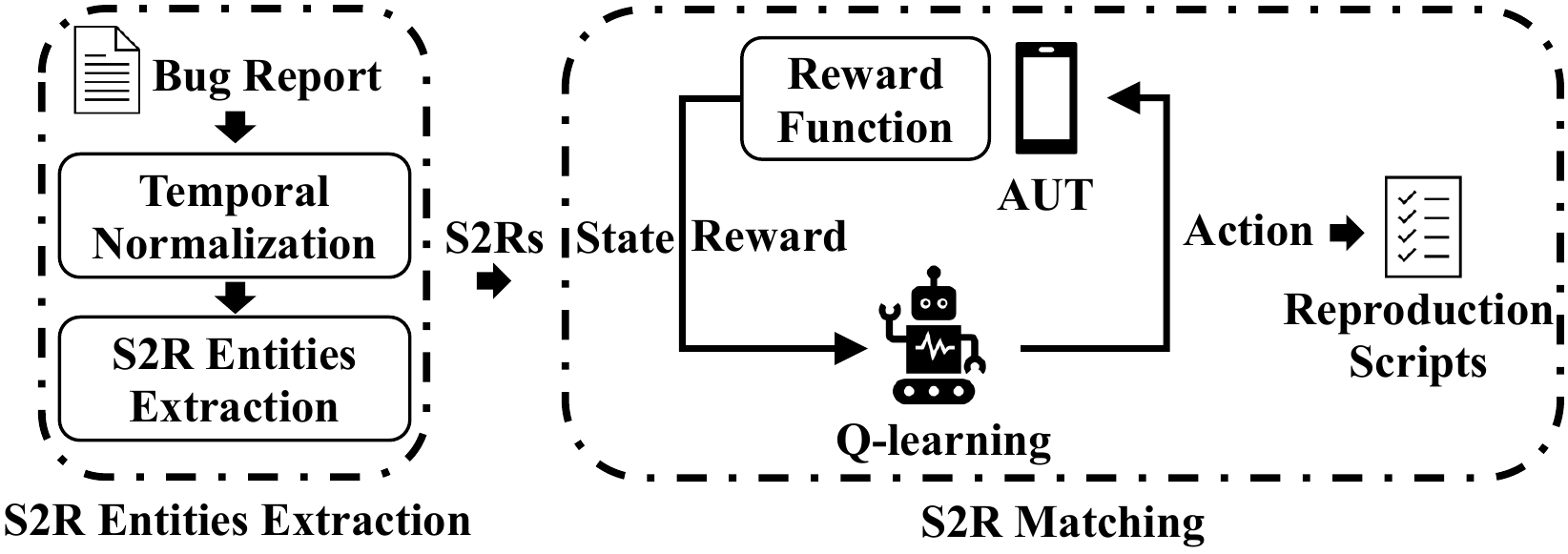}
\caption{Workflow of Our Approach}
\label{fig:workflow}
\end{figure}

%% file: content/approach_s2r_extraction.tex
\subsection{Extraction of S2R Entities}
\label{sec:approach:s2r_extraction}
\input{content/pictures/temporal_reorder}

The first phase of our approach analyzes the text of a bug report and extracts the individual steps defined in each sentence of the report.  The input to this phase is the list of sentences in the bug report.  For each of these sentences our approach first carries out a temporal normalization analysis that extracts the \acp{s2r} in the sentence, converts them to standalone sentences, and then reorders them based on the implied temporal relationships in the original sentence (\Cref{sec:approach:identify_temporal_relation}). The ordered standalone sentences are then each analyzed to infer \acp{s2r} entities --- important reproduction information, such as the target widget, action type, and input values (\Cref{sec:approach:s2r_entities}).  Our \ac{s2r} entity inference is notable from related work since it does not require the \acp{s2r} to match one of a predetermined set of patterns.  The final output of the first phase is a list of \acp{s2r} entities, each corresponding to a reproduction step that the second phase of our approach (\Cref{sec:approach:reproduction}) will use as it attempts to reproduce the bug report.

\subsubsection{Normalizing the Temporal Order of S2Rs}
\label{sec:approach:identify_temporal_relation}

The first step of the extraction analysis normalizes the temporal order of reproduction steps described in a bug report sentence. For example, this step would address the problem shown in the text of \Cref{sec:motivating_example}, where the bug report sentence has multiple conjuncted \acp{s2r} whose intended execution order is not the same as their syntactic ordering.  Reasoning about the temporal relations among \acp{s2r} is challenging due to the use of complex sentence structures, such as nested clauses and phrases. Our insight is that this temporal ordering information can be extracted by accounting for the semantics of connectives utilized between clauses or phrases in a bug report sentence. To take advantage of this insight, we designed an analysis that, given an \ac{s2r} sentence, recursively: (1) extracts the conjuncted \ac{s2r} text spans (\ie clauses or phrases) and connectives, (2) converts each \ac{s2r} text span into a standalone sentence, and then (3) reorders the standalone sentences based on the connectives between them. Each round of our analysis generates a pair of ordered standalone sentences. Our analysis is recursively performed on the generated sentences until no more conjuncted \acp{s2r} can be identified.
The output of our analysis is a list of the standalone sentences in the inferred temporal order. We next describe the details of each part of our analysis in detail, illustrating the analysis using the sentence in \Cref{fig:bugreport}.

As a preprocessing step, each round of our analysis takes a given sentence and converts it into a standard structure for representing a natural language sentence.
This structure is known as the \ac{cpt} and captures the syntax of a sentence from the perspective of constituency grammars~\cite{speechAndLanguageProcesssing}.  \Cref{fig:cpt} shows the \ac{cpt} of the sentence in \Cref{sec:motivating_example}, which is computed using standard NLP parsers~\cite{chenFastAccurateDependency2014}. Each terminal node of the \ac{cpt} denotes a word in the sentence and each non-terminal node represents a type of constituency, which captures the use of all terminal nodes in its sub-tree as a single unit~\cite{pennTreeback} (\eg \textit{NP} for a noun phrase).

Given the \ac{cpt}, our analysis first extracts the conjuncted \ac{s2r} text spans (\ie clauses and phrases) and their connectives in the sentence.  
In order to extract the text spans, our analysis traverses each constituency tag in the \ac{cpt} from top to bottom and identifies two types of conjunction: coordination and subordination, as defined using standard English grammar rules~\cite{speechAndLanguageProcesssing}.
To identify coordination, our approach searches for tags whose parent-child structure satisfies the coordination grammar~\cite{speechAndLanguageProcesssing}.
For example, the grammar for coordinated clauses is defined as "$S\rightarrow S\ CC\ S$", which indicates that a parent constituency $S$ (sentence) has three children: two sub-clauses and a connective $CC$ between them.
By identifying such a parent constituency $S$, our analysis can find the two coordinated sub-clauses and their connectives from its children.
To identify subordination, our approach searches for its corresponding constituency tag, $SBAR$, which represents subordinating clause, and takes the words rooted at this tag as the text for a subordinate clause and the first word in the clause as the connective.
This step of our analysis returns the first pair of conjuncted text spans it identifies as it traverses the \ac{cpt} from top to bottom. This ensures that every round our analysis identifies the most outer conjuncted text spans. 
For example, given the \ac{cpt} in \Cref{fig:cpt}, our approach identifies two coordinated verbal phrases (VP) as highlighted and the word "and" as the connective in the first round of analysis. 

The next step of our analysis transforms each text span identified by the previous step into a standalone sentence.
This transformation is necessary since part of the text in the original sentence may be shared by the conjuncted texts (\eg the subject "I" is shared by all the four \acp{s2r} in the sentence in \Cref{fig:bugreport}). Only extracting and reordering the conjuncted text will lose such information.
To do this, our approach first removes the conjuncted constituency tags (identified in the previous step) and their sub-tree from the original \ac{cpt}. This gives us the part of the original \ac{cpt} that is shared by the conjuncted text spans.
Then our approach duplicates this part and joins it to the subtree of each conjuncted text to form new \acp{cpt}, representing the transformed sentences. To illustrate this transformation, after identifying the two highlighted verbal phrases in \Cref{fig:cpt}, our approach extracts the left subtree of the root node, which indicates the subject "I", duplicates it, and rejoins it to the subtree of each verbal phrase. This forms two standalone sentences. 

The last part of our analysis infers the intended execution order of the transformed standalone sentences. 
Our inference is based on the semantics of the connectives between the sentences. To obtain a comprehensive view of connectives used in English text and their semantics, we referred to the \ac{pdtb}~\cite{miltsakakiPennDiscourseTreebank2004,prasadPennDiscourseTreeBank2008}.
The \ac{pdtb} contains a large-scale annotation on the English connectives and provides a comprehensive categorization based on their semantics. Our approach focuses on the connectives with semantics in two categories from \ac{pdtb}: \textit{temporal succession} and \textit{alternative}. Connectives in the first category indicate that the second conjuncted text span happens earlier than the first conjuncted span, and connectives in the second category indicate that the conjuncted text spans could be used alternatively to represent the meaning of the sentence. The reason to focus on these two categories is that connectives within them would affect the temporal order or the selection of \acp{s2r}. 
To perform the temporal inference, given the connective, our approach first checks which category it belongs to. If the connective belongs to the first category, our approach reverses the order of conjuncted sentences. If the connective belongs to the second category, our approach only selects one from the given sentences.
For the example in \Cref{fig:bugreport}, our approach would reverse the order of standalone sentences generated by the first two \ac{s2r} text spans as it finds the connective "when" indicates a temporal succession semantics.
\Cref{fig:reorderedsentence} displays the final sentence list generated by our tool where each sentence only contains one reproduction step and their ordering is normalized.

\subsubsection{Inferring \ac{s2r} Entities from Standalone Sentences}
\label{sec:approach:s2r_entities}

The goal of this part of the approach is to infer \ac{s2r} entities from the standalone sentences produced by the analysis in \Cref{sec:approach:identify_temporal_relation}.  These \ac{s2r} entities will be used in the exploration phase of the approach (\ie \Cref{sec:approach:reproduction})  to navigate the \ac{ui} of the \ac{aut}.  We formally define the entities of a \ac{s2r} as $\langle$target widget, \ac{ui} action, input value, target direction$\rangle$. Taken together, these elements represent the \ac{ui} event described in the standalone sentence. Prior work~\cite{zhaoReCDroidAutomaticallyReproducing2019,fazziniAutomaticallyTranslatingBug2018} targeted the identification of a similar set of \ac{s2r} entities. However, as we discussed in \Cref{sec:motivating_example}, these approaches can only work for predefined sentence patterns since they require a mapping of parts of the pattern to the \ac{s2r} entities.   Our insight is that each of these \ac{s2r} entities can instead be inferred based on more general syntactic constituents (\ie subject, predicate, object, and modifier) of a sentence. For example, the predicate of a sentence, regardless of where it appears in a given sentence, can be used to infer the action performed by the user.  This insight allows our approach of extracting \ac{s2r} entities to work on any standalone sentence, regardless of its form, as long as it can be decomposed into these syntactic constituents --- a condition that would be satisfied by any standalone sentence produced by our analysis in \Cref{sec:approach:identify_temporal_relation}. Our approach leverages this insight as follows. First, given a standalone sentence produced by the analysis in \Cref{sec:approach:identify_temporal_relation}, our approach decomposes the sentence into its syntactic constituents using standard \acl{openie} techniques~\cite{etzioniOpenInformationExtraction2008a,wuOpenInformationExtraction2010,delcorroClausIEClausebasedOpen2013a,schmitzOpenLanguageLearninga,bhutaniNestedPropositionsOpen2016,cettoGrapheneSemanticallyLinkedPropositions2018}.  In the second step, our approach identifies the text in the sentence that defines the \ac{s2r} entities by using inference rules based on these syntactic constituencies.  In the remainder of this section, we first give an example of the syntactic constituents that our approach works on and then define the inference rules we use for each of the parts of the \ac{s2r} entity.

For background purposes, we provide an example of the syntactic constituents using the last sentence in \Cref{fig:reorderedsentence}. The predicate of this sentence is "click," the subject is "I", the object is "the CANCEL button," and the modifier is the phrase "in the circle". Notably, the modifier is optional for a sentence. To retrieve all these constituents, our approach in practice employs a recent \ac{openie} technique~\cite{cettoGrapheneSemanticallyLinkedPropositions2018}. Next we explain the rules we defined to infer each \ac{s2r} entity.

The \textbf{target widget} is the text description of the UI widget that the user interacts with in the \ac{s2r}.  The part of the sentence that contains the target widget depends on the voice with which the sentence was written.  If the sentence is written in an active voice, the target widget is defined in the sentence object (\eg "I click \underline{the button}").  Otherwise, if the sentence is written in the passive voice, the target widget is defined in the subject of the sentence (\eg "\underline{The button} is clicked").  Therefore, to identify the target widget, our approach first infers the voice of the sentence.  
The voice can be determined simply by checking whether the predicate is in a form of a static verb and a verb in its past participle (\eg is clicked)~\cite{pastvoicewiki}.
Based on this determination, our approach extracts either the subject or object text, which is part of the syntactic constituents defined previously and then uses this text for the target widget.

The \textbf{\ac{ui} action} represents the type of interaction that the sentence describes as performed on the \ac{aut}. This naturally corresponds to the predicate part of the sentence, so our approach maps the predicate to one of the five standard actions that can be performed on an Android \ac{ui}: click, input, rotate, swipe, and scroll.   However, the key challenge is that actions may be described using a wide range of words, so there is not a direct mapping from the predicate to the known action types.  Existing approaches rely on a predetermined list of synonyms for each known action type to identify a direct mapping.  The limitation of this approach is that if the predicate uses previously unseen words, it cannot be classified.  Our insight is to classify the \ac{ui} action type of a given predicate based on its semantic similarity with a set of synonyms defined for each standard \ac{ui} action. 

Our approach for action type identification is as follows.  We assume the availability of a word list, as used in related approaches (\eg \cite{zhaoReCDroidAutomaticallyReproducing2019}), that contains a group of synonyms for each standard action type.  For example, "press" and "tap" for the click action.  To handle action types described using words that are not in the list, we introduce the idea of semantic similarity to the action type identification.  Our approach computes the semantic similarity of the predicate with a group representing the action type and its synonyms. The semantic similarity enables our approach to know which type of standard \ac{ui} action the predicate is most similar to semantically.  
Following the classic method~\cite{mikolovDistributedRepresentationsWords2013}, given two words or phrases, we compute their semantic similarity as the cosine distance of their word embeddings, the vector representation of a text, obtained from a pre-trained language model, such as Word2Vec~\cite{mikolovDistributedRepresentationsWords2013}.
Our approach considers the group with the highest semantic similarity to be the inferred action type. However, if the similarity score difference of the top two groups is within a threshold $\delta$ then both of the corresponding action types are considered during the reproduction phase (\ie \Cref{sec:approach:reproduction}).  This aspect of our approach allows it to more flexibly handle words that could be interpreted in multiple ways. For example, the word "change" could be interpreted as a click action or an input action in different cases.  If the inferred action type is rotate, our approach also checks whether its inferred target widget describes the screen or the device. If neither is described, our approach chooses the next highest scored \ac{ui} action as the result. The reason for this is that intuitively the rotation action is performed on the whole device or the screen instead of a specific widget, so if neither is the target widget, then the action is not rotate. 
This heuristic helps our approach identify rotation action more accurately.

The \textbf{input value} is the text to be entered for an input action.  This information can be inferred by analyzing the object and the modifier part of a sentence. Due to the way actions are described in the English language, input values are associated with their targets using prepositional phrases. This is captured by the modifier of a sentence. However, depending on the preposition used in the modifier, the input value may be contained in different constituents of the sentence. For example, for the sentence ``I enter A on B'', the input value \textit{A} is captured by the object. However, for the sentence ``I fill A with B'', the input value \textit{B} is contained in the modifier. Therefore, our approach identifies the input value based on the preposition used in the modifier. If the modifier starts with the preposition "with", our approach extracts the text following it as the input value. If the modifier starts with a preposition in \{"in", "on", "into", "onto", "at"\}, our approach uses the object as the input value. It also replaces the previously identified target widget with the modifier text. In the case that there is no modifier identified in the given sentence, which happens when reporters only specified the input value or target widget as the object of the sentence (\eg "I enter \underline{a number 13}"), our approach takes the object text as the input value. Our approach employs two heuristics to refine the identified input value text to make it more precise. First, if the input value text contains numbers, our approach only keeps the number as the input value. The reason for this heuristic is that reporters may include other words or phrases when describing the input number as the underlined phrase in the previous example. However, the input value to reproduce the failure needs to be very precise, in this case, only the number is entered. Second, if the text uses a text-based description of a special value, such as "space", our approach replaces it with the corresponding literal value.

The \textbf{target direction} defines the intended direction for a scrolling action (\ie up or down) or a swiping action (\ie left or right). The direction is also expressed in the object portion of a sentence. For example, "I scroll \underline{down}". Given the simplistic nature of this entity, we found it sufficient for our approach to directly search for directional keywords in the object text and use those as the target direction.

%% file: content/pictures/temporal_reorder.tex
\begin{figure}[t]
  \centering
    \begin{subfigure}[t]{0.35\textwidth}
        \centering
        \includegraphics[width=\textwidth]{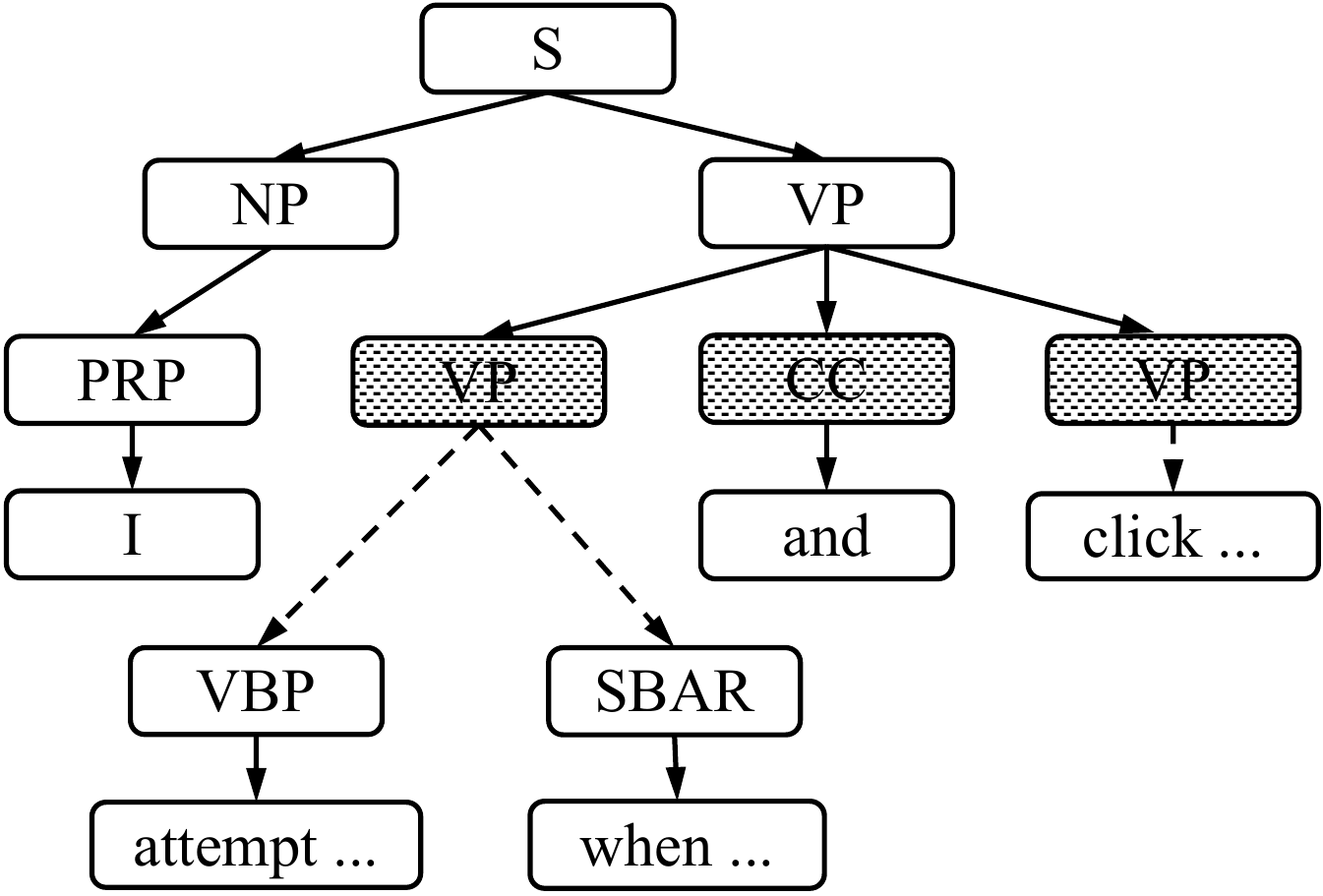}
        \caption{Constituency Parsing Tree}
        \label{fig:cpt}
    \end{subfigure}
    \begin{subfigure}[t]{0.35\textwidth}
        \centering
        \includegraphics[width=\textwidth]{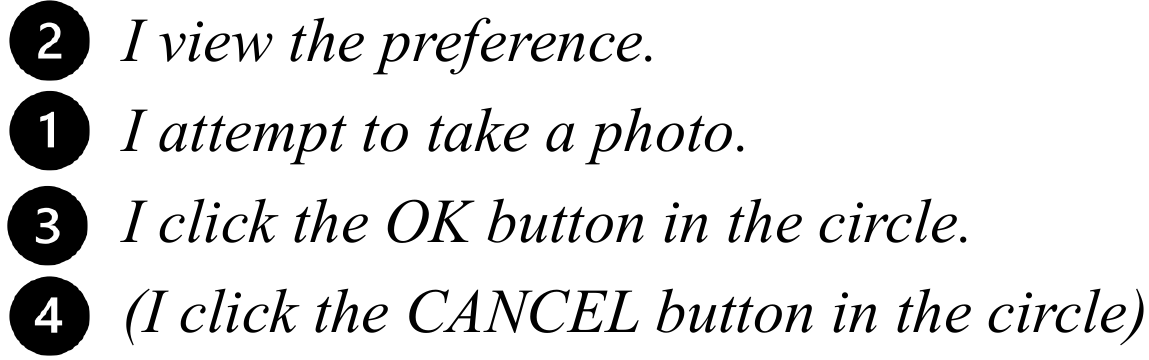}
        \caption{Reordered Sentences}
        \label{fig:reorderedsentence}
    \end{subfigure}
\caption{Example of \acp{s2r} Reordering}
\label{fig:clausalphrsalexample}
\end{figure}

%% file: content/approach_s2r_matching.tex
\subsection{Matching \acp{s2r} to \ac{ui} Events}
\label{sec:approach:reproduction}

During the second phase, \ac{s2r} matching, our approach explores the \ac{aut} to match the \acp{s2r} to a sequence of \ac{ui} events that reproduce the failure.  As explained in \Cref{sec:motivating_example}, for state of the art techniques, their exploration-based process of finding such \ac{ui} events is limited due to the problems of local optimums and missing steps. Our approach addresses these challenges by defining the matching problem as a \acf{mdp} and using a type of reinforcement learning algorithm, \qlearning, to find the desired \ac{ui} event sequence.

The combination of \ac{mdp} and \qlearning is well-suited for our problem domain and addressing these challenges. The expected reward used by \qlearning combines the immediate reward for the current action and the accumulated reward of potential future actions. This enables our approach to avoid local optimums by allowing it to consider and evaluate a \ac{ui} event that itself may not be the closest match with the \acp{s2r} but leads to subsequently better matched \ac{ui} events. For missing steps, a similar mechanism can be leveraged. In this case, a \ac{noop} is added to the actions in the \ac{mdp} for a given state. When combined with the \qlearning future rewards, this effectively allows our approach to evaluate the total expected rewards if it assumes the next step is missing.
The \ac{noop} action can be added repeatedly to simulate the possibility of multiple missing steps. Taken together, the combination of \qlearning and \ac{mdp} enables our approach to break out the local optimum and bridge missing steps effectively.

In the following sections, we first provide a formal definition of our instance of the \ac{mdp}, which maps it to the problem of \ac{s2r} matching (\Cref{sec:mdp}).  Next, we explain how our approach utilizes \qlearning on this \ac{mdp} to find a \ac{ui} event sequence that reproduces the bug report (\Cref{sec:matching}).

\subsubsection{Formulation of Markov Decision Process}\label{sec:mdp}
In our approach, we define an instance of the \ac{mdp} to describe the process of matching \acp{s2r} with \ac{ui} events in the \ac{aut}.  The \ac{mdp} is formally defined as a tuple $\langle \mathcal{S}, \mathcal{A}, \mathcal{P}, \mathcal{R} \rangle$ where $\mathcal{S}$ is the state set, $\mathcal{A}$ is the action set, $\mathcal{P}$ is the transition function, and $\mathcal{R}$ is the reward function. We define each part of the \ac{mdp} for our problem as follows:

\textbf{$\mathcal{S}$: \textit{States}}
The state represents the \ac{mdp} at each time step.
We define a state of the \ac{mdp} as $s = \langle H, RS, n \rangle$. 
The first element $H$ is the current \ac{vh} of the \ac{aut}, which includes all the widgets on the \ac{ui} as well as their attributes.
The second element $RS$ is a list of the remaining \acp{s2r} that are not yet matched.
The last element $n$ is an integer indicating the remaining amount of \ac{noop} actions that are available for the matching process. Our approach defines a finite limit on \acp{noop} actions to avoid the situation where the reproduction phase can endlessly explore an app along all paths with no likely  matches. Allowing an infinite number of missing steps would be unrealistic, implying a bug report that had an infinite (or extremely large number) of missing steps. We define the initial state as $s_0=\langle H_0, RS_0, n_0 \rangle$ where $H_0$ is the \ac{vh} of the \ac{ui} where the reproduction starts, $RS_0$ is a list containing all of the \acp{s2r} in the bug report identified by the first stage analysis of our approach and $n_0$ is the total number of allowed \ac{noop} actions during whole matching process. We define the terminal state of the \ac{mdp} as $s_t$, which has no available actions, \ie it is the ending of the matching process. This happens when our approach matches all \acp{s2r} and used all available \ac{noop} actions.


\textbf{$\mathcal{A}$: \textit{Actions}}
In the context of \ac{s2r} matching problem, the action represents a matching between a \ac{ui} event and an \ac{s2r}. 
(Note, this terminology can be confusing since in \Cref{sec:approach:s2r_entities}, action refers to the standard Android operations, such as click and swipe.)   
We formally represent an action $a$ as $\langle e, rs \rangle$ where $e$ is the \ac{ui} event that is chosen to be executed on the \ac{aut} and $rs$ is the step matched with the \ac{ui} event.
To handle missing steps, we define a special action, \ac{noop}, which is denoted as $\langle e, - \rangle$ where the dash represents a placeholder for a missing step.
We will discuss how our approach identifies available actions at a given state in \Cref{sec:matching}.


\textbf{$\mathcal{P}$: \textit{Transition Function}}
The transition function of an \ac{mdp} takes in the current state $s$ and the selected action $a$ and returns the next state $s'$. 
In our problem domain, the transition to next state happens in two steps.
First, our approach executes the \ac{ui} event $a.e$ on the \ac{aut} and extracts the new \ac{vh} produced by the app. Then, our approach dequeues the selected step from $s.RS$ or decreases $s.n$ if $a$ is a \ac{noop} action.  The new \ac{vh} and the possibly updated values of $RS$ and $n$ are returned as the new state.

\textbf{$\mathcal{R}$: \textit{Reward Function}}
The reward function generates a value indicating the quality of the action, \ie the quality of the match between an \ac{s2r} and a \ac{ui} action. Our reward function evaluates an action as a sum of three subscores: similarity score, exploration penalty, and failure state penalty. 

The \textit{similarity score} measures the similarity between an action's \ac{ui} event and \ac{s2r}.
It guides our approach to explore \ac{ui} events closely related to the descriptions in the bug report.
The computation of similarity scores is done in two ways.
First, for actions matching with \ac{ui} events interacting with a specific widget, \ie click and input events, the similarity score is computed as the textual similarity between the description of the widget on the \ac{ui} and the target widget entity of the \ac{s2r}.
This textual similarity indicates how related the \ac{ui} event is to the \ac{s2r}.
To do this, our approach analyzes the \ac{vh} of the \ac{ui} and extracts three attributes of the widget as its description:
\begin{enumerate*}
    \item \textit{text}: This attribute is the text appearing on the widget, which is readily available to bug reporters;
    \item \textit{id}: This attribute contains the file name of the linked resources used in the widget, such as icons. The file name is normally meaningful and descriptive of the content it contains~\cite{fazziniAutomaticallyTranslatingBug2018}; and
    \item \textit{content description}: This is the description of the widget defined by the developers. Its content is used by Android accessibility service (\eg screen reader) when describing a widget to people with disabilities~\cite{contentDescription}. Therefore it is supposed to contain a meaningful and informative explanation of the widget. 
\end{enumerate*}
Our approach computes the semantic similarity of the text in the target widget entity with each of these three sources of description text, using the same approach described in \Cref{sec:approach:s2r_extraction}.  Our approach uses the highest of these three values as the overall similarity score, as long as the value is above a threshold \simthred that represents the similarity of non-synonym words~\cite{zhaoReCDroidAutomaticallyReproducing2019}.  However, if the value is not above this threshold then our approach considers them to not be synonyms and assigns a default negative score \defaultsimscore. The rationale for using the maximum of the three values is that our approach cannot know a priori which of these three descriptions sources would be used by the bug reporter, so this mechanism allows for using the best or most informative fit.
Second, for actions that match with rotate, scroll, swipe, or the \ac{noop}, our approach assigns their similarity score as the default score \defaultsimscore. The reason for this is that for these actions, the matched \ac{ui} event does not interact with a specific \ac{ui} widget or the matched \ac{s2r} is not a concrete step from the bug report. Therefore, our approach cannot compute a score to show the relevance or similarity for the \ac{ui} event and the \ac{s2r}.


The \textit{exploration penalty} \explorationpenalty is generated when the selected \ac{ui} event does not change the \ac{ui} of the \ac{aut}. The intention of this penalty is to guide our approach to avoid matching \acp{s2r} with such meaningless \ac{ui} events. For example, in the \ac{ui} with a tab, it is meaningless to click the current tab button again. By adding a penalty to these cases, our approach focuses on matching meaningful \ac{ui} events. To detect such cases, our approach measures whether the \ac{vh} of the \ac{ui} is changed before and after executing the selected \ac{ui} event. If all the widgets and their status on the \ac{vh} remain the same, the penalty is generated for the action.


The \textit{failure state penalty} \failurepenalty is assessed when the selected action leads to a terminal state that does not reproduce the observed failure. This penalty discourages the exploration of event sequences that do not lead to the observed failure.


\subsubsection{Match \acp{s2r} to \ac{ui} events}\label{sec:matching}

\input{content/algorithms/s2r_matching_algo}

This part of our approach utilizes \qlearning to find a \ac{ui} event sequence that matches the given \acp{s2r} and reproduces the observed failure of the bug report. Our approach is described in \Cref{alg:s2r-matching}. As input, our approach takes the \acf{aut}, \acfp{s2r}, a time budget (t), and the error message (o) of the observed failure as inputs.  Our approach then iteratively explores the \ac{aut}, guided by the \qlearning, to find a \ac{ui} event sequence that reproduces the failure.

The iterative exploration starts from the initial state $s_0$ (\Cref{line:s2r-matching:resetstate}). 
At each iteration, our approach first identifies all the possible actions ( $\mathcal{A}$), available in the current state (\Cref{line:s2r-matching:inferaction}).
Recall that an action in the \ac{mdp} is actually a possible match between an \ac{s2r} and a \ac{ui} event.  
Our approach identifies the set of possible matches from two sources.  
First, the match set includes the cross product of all available \ac{ui} events in the \ac{ui} and the next \ac{s2r} in $s_i.RS$, if the two represent the same \ac{ui} action. 
In the case that our approach predicted two possible types of \ac{ui} actions for a \ac{s2r}, it includes matches for both of them. 
Second, the match set also includes a possible \acp{noop} match for all \ac{ui} events, if the \ac{noop} limits has not been reached at $s_i$ (\ie $s_i.n > 0)$. 

After identifying the set of all possible matches, our approach selects one match and performs the matched \ac{ui} event on the \ac{aut}. Our approach selects the match using the epsilon-greedy algorithm~\cite{reinforcementlearning}, a standard method for \qlearning. Specifically, with probability $1-\epsilon$, our approach will choose the match with the highest \qvalue (\Cref{line:s2r-matching:exploitation}) or with probability $\epsilon$, select a random one from the matches with lower \qvalue (\Cref{line:s2r-matching:exploration}). 
In practice, our approach sets the $\epsilon$ with a low value so that our approach can focus on exploring the current optimal choice (\ie the match with highest \qvalue) in most time in order to match more \acp{s2r} along this path, but also have chance to explore other random actions to break out the local optimum. 
It is worth emphasizing that, the epsilon-greedy policy gives our approach a chance, at any given state, to explore lower-scored matches which could be a match between a \ac{ui} event that is not the most similar to the \ac{s2r} or a match with a missing step holder. By doing this, our approach could evaluate their future rewards, which enables it to effectively escape the local optimal during the exploration and, in combination with the \ac{noop}, bridge missing steps.
In the case that a previously unexplored state is matched, the initial \qvalue for all the actions is set to its \textit{similarity score} as defined in \Cref{sec:mdp}. 
By doing this, our approach is given a pre-knowledge of the potential reward of each action so it can be more effective selecting a better match with less exploration. For the  scroll, swipe, or rotate actions, since a similarity score is not defined, the approach assumes a similarity threshold $t$ (defined in \Cref{sec:mdp}) as their initial \qvalue. This adjustment ensures that these impactful actions will be explored.
The selected \ac{ui} event is stored in an event list, which is returned in the case of successful reproduction (\Cref{line:s2r-matching:addevent}). By calling the transition function, our approach executes the selected \ac{ui} event on the \ac{aut} and updates the current state (\Cref{line:s2r-matching:statetransition}).

After executing the \ac{ui} event in the selected match, our approach then determines its reward (\Cref{sec:approach:identify_temporal_relation}) and updates its \qvalue using the Bellman function (\Cref{line:s2r-matching:updateqvalue}). In the case of encountering a terminal state (\Crefrange{line:s2r-matching:checkterminate}{line:s2r-matching:terminateepoch}), our approach resets the current state and restarts the exploration. Our approach terminates either when the timeout occurs or the failure is successfully reproduced (\Cref{line:s2r-matching:return}). The success of the reproduction is determined by checking whether the \ac{aut} crashes with the specified error message (\Cref{line:s2r-matching:checksuccess}).

%% file: content/algorithms/s2r_matching_algo.tex
\RestyleAlgo{ruled}
\LinesNumbered
\begin{algorithm}[t]
\caption{\ac{s2r} Matching}
\label{alg:s2r-matching}
\KwIn{the app under test $AUT$, reproduction steps $S2Rs$, timeout $t$, error message $m$}
\KwOut{reproduction \ac{ui} events}
$Q \gets \emptyset$ \tcc*{initialize Q-table} \label{line:s2r-matching:initqtable}
\While{$\neg timeout(t)$}{
    $s_{i+1} \gets s_0$ \tcc*{reset to initial state} \label{line:s2r-matching:resetstate}
    $E \gets \emptyset$\; \label{line:s2r-matching:reseteventlist}
    \While{true}{
        $s_i \gets s_{i+1}$\;
        $\mathcal{A} \gets \text{inferActions}(s_i)$\; \label{line:s2r-matching:inferaction} 
        \uIf{$\text{randomNum}() < \epsilon$}{ \label{line:s2r-matching:actionselection1}
            $a_i \gets \text{rand}(\{ a\in \mathcal{A}\mid a \neq arg\,max_{a'}Q(s,a')\})$\; \label{line:s2r-matching:exploration} 
        }
        \Else{
            $a_i \gets arg\,max_{a}Q(s,a)$\; \label{line:s2r-matching:exploitation} 
        } \label{line:s2r-matching:actionselection2}
        $E\gets E \cup a_i.e$\;\label{line:s2r-matching:addevent}
        $s_{i+1} \gets \mathcal{P}(a_i, s_0, AUT)$\;\label{line:s2r-matching:statetransition}
        \If{success($md$)}{ \label{line:s2r-matching:checksuccess}
            \Return{E}\; \label{line:s2r-matching:return}
        }
        $r \gets \mathcal{R}(a_i, s_i, s_{i+1})$\;\label{line:s2r-matching:computereward} 
        $Q(s_i, a_i) \gets (1-\alpha)Q(s_i,a_i) + \alpha (r +\gamma Q^*(s_{i+1},a))$\;\label{line:s2r-matching:updateqvalue} 
        \If{$s_{i+1}$ is a terminal state}{ \label{line:s2r-matching:checkterminate}
            \Break\; \label{line:s2r-matching:terminateepoch}
        }
    }
}
\end{algorithm}

%% file: content/evaluation.tex
\section{Evaluation}


In this section, we evaluated the effectiveness and runtime of our approach.  We also compared our approach against two state of the art bug reproduction approaches, \recdroid and \yakusu. Our evaluation addressed the following research questions:

\noindent
\textbf{RQ1}
How effective is our approach in identifying \acp{s2r}?\\
\textbf{RQ2}
How effective is our approach in reproducing Android bug reports?\\
\textbf{RQ2.1}
How helpful is the \ac{nlp} techniques for reproduction?\\
\textbf{RQ2.2}
How helpful is the \acs{rl}-based exploration for reproduction?\\
\textbf{RQ3}
What is the running time of our approach?

\subsection{Approach Implementations}
\label{sec:implementation}

For the purposes of the evaluation, we implemented a prototype, \tool, of our approach. The core algorithms of our approach are implemented in Python, and we leveraged functionality from several well-known libraries: Graphene~\cite{cettoGrapheneSemanticallyLinkedPropositions2018} and OpenIE5~\cite{openIEGH} to identify syntactic constituents in natural language text;  UIAutomator~\cite{uiautomator} to interact with Android apps, and Spacy~\cite{SpacyPaper} to compute the semantic similarity of texts.  Our implementation will be publicly available via our group's GitHub repository, but is currently blinded for review~\cite{toolGH}. We obtained the implementations of \recdroid~\cite{zhaoReCDroidAutomaticallyReproducing2019} and \yakusu~\cite{fazziniAutomaticallyTranslatingBug2018}, two state of the art reproduction tools on Android bug reports, from their public websites\cite{recdroidGH,yakusuWebSite}. Our experiments were performed on Android emulators on a physical x86 Ubuntu $20.04$ machine with eight 3.6GHz CPUs and $32$G of memory.


\subsection{Dataset Collection}\label{sec:eval-dataset}

To evaluate our tool, we built a dataset containing $77$ bug reports. We collected the bug reports from the artifacts of four related works: (1) the evaluation dataset of \recdroid ~\cite{zhaoReCDroidAutomaticallyReproducing2019,zhaoReCDroidAutomatedEndtoEnd2022}; (2) the evaluation dataset of \yakusu~\cite{fazziniAutomaticallyTranslatingBug2018}; (3) an empirical study on Android bug report reproduction~\cite{johnsonEmpiricalInvestigationReproduction2022}, and (4) an Android bug report dataset~\cite{wendlandAndror2DatasetManuallyReproduced2021}. Altogether, there were $199$ bug reports from these four sources. For each bug report, the authors interacted with the corresponding app to both ensure that the selected bug report was reproducible and to remove any duplicates. As part of this process, we also separately documented the steps that were missing from the bug report but that were necessary to reproduce the reported failure. During this process, a total of $99$ bug reports were found to be no longer reproducible and $23$ were duplicates, leaving us with $77$ bug reports. The lack of reproducibility of a given bug report was generally attributable to one or more of the following reasons: the lack of available apks or bug reports (due to an expired link), environmental configuration issues that were not resolvable, or failure to obtain the necessary setup information for the app (e.g., personal accounts or specific inputs). Additionally, we manually analyzed and identified the ground truth for each step within the $77$ bug reports, specifically the individual \acp{s2r} contained within each step and each individual \ac{s2r}'s \ac{ui} action, target widget, input value, and target direction. The average, median, minimum and maximum steps of the ground truth of the bug reports were: $3$, $3$, $1$, and $9$. However, after manually reproducing each bug report and identifying the actual number of required steps for each bug report reproduction, we found that the average number of required steps increased to $6$, the median to $5$, and the maximum to $26$, while the minimum remained at $1$. 
Additional details regarding the subject apps and bug reports in this study are available in our supplementary data, which will also be made available via the project website.


\subsection{Experiment 1}
\subsubsection{Protocol} 

The first experiment evaluated the accuracy of \tool in identifying \acp{s2r} and compared these results with those of \recdroid and \yakusu. To do this, we ran \tool, \recdroid, and \yakusu on our dataset and verified the correctness of all identified \acp{s2r} against the ground truth. An \ac{s2r} extracted by a tool was determined to be correct if all of the following conditions were satisfied:
\begin{enumerate*}
    \item the action type was an exact match with the action type from the ground truth;
    \item the target widget, if it existed, included target widget words from the ground truth;
    \item the input value, if it existed, was an exact match with that from the ground truth; and
    \item the target direction, if it existed, was an exact match with that from the ground truth.
\end{enumerate*} 
As metrics for the accuracy of the three tools, we calculated precision by dividing the number of correctly extracted \acp{s2r} by the total number of extracted \acp{s2r}, and calculated recall by dividing the number of correctly extracted \acp{s2r} by the total number of \acp{s2r} in the ground truth.

\subsubsection{RQ1 \ac{s2r} Entity Extraction Results}\label{sec:rq1_result}

Our results were as follows: \tool extracted a total of $294$ \acp{s2r} and $198$ correct \acp{s2r}, and it achieved a precision of $67\%$ and a recall of $77\%$. \recdroid extracted a total of $220$ \acp{s2r} and $116$ correct \acp{s2r}, and it achieved a precision of $53\%$ and a recall of $45\%$. Lastly, \yakusu extracted a total of $260$ \acp{s2r} and $147$ correct \acp{s2r}, and it achieved a precision of $57\%$ and a recall of $59\%$.  These results show that \tool was able to effectively identify \acp{s2r} and was more accurate than \recdroid and \yakusu for extracting \acp{s2r}.

We also investigated the specific reasons for our tool’s inaccurate \acp{s2r} and found five categories of root causes:
\begin{enumerate*}
    \item The \ac{ui} action word list used by \tool was not comprehensive.  For example, because \tool was unable to find a semantically similar word in the word list for the click action, our approach misclassified the word "navigate" as an input action. Our approach was unable to accurately classify the \ac{ui} action of a given predicate word in these cases.
    \item The imprecision of the underlying \ac{nlp} techniques was also found to be a cause of failure in our tool. Specifically, the underlying techniques occasionally failed to identify the syntactic constituents accurately, which caused our tool to misidentify the target widget. Additionally, the underlying tool sometimes failed to identify conjunction relations between clauses, which caused our tool to miss the \acp{s2r} in those clauses. 
    \item A few \acp{s2r} were found to have target widgets described via the predicate of the sentence. In these cases, our tool failed to extract the target widgets. An example of this is ``I finish the dialog'', where it expresses a step to press the ``finish'' button in the \ac{aut}. \tool was unable to extract the target widget in this case since it assumes the predicate is used for describing the \ac{ui} action.
    \item Bug reports were occasionally found to have utilized a special or unique way to express required input values to trigger a crash. Our tool relies on bug reports expressing input values in the specific literal form for input, rather than using descriptive language, and, thus, failed to extract input values correctly when these situations occurred. For example, one bug report included ``enter a number larger than Integer.MAX\_VALUE''.  \tool did not identify the correct input value because it was not directly specified in the bug report.
    \item A few bug reports contained text that either did not relate to or was not required to reproduce the described crash. In these situations, our tool often extracted extra \acp{s2r} from such sentences, which were not intended to be a part of the desired reproduction steps. Our approach assumes that all text contained within a given bug report is intended to be included in the reproduction process, which can lead to failure if this was not the case. Our tool is unable to identify sentences that should not be considered a part of the reproduction process.
\end{enumerate*}

\subsection{Experiment 2}
\subsubsection{Protocol}

The second experiment evaluated the effectiveness (RQ2) and running time (RQ3) of our approach. To carry out this experiment, we ran \tool on the collected dataset, and then compared its results against those of \recdroid and \yakusu.  In addition to RQ2, we introduced two sub RQs that evaluated the contributions of the two parts of our approach independently. RQ2.1 evaluated the \ac{s2r} extraction (\Cref{sec:approach:s2r_extraction}). For this RQ, we created a variant of our approach, \toolvara, that did not use the \ac{s2r} extraction method described in \Cref{sec:approach:s2r_extraction}, and instead, utilized the \acp{s2r} extraction method from \recdroid to drive the reproduction phase. RQ2.2 evaluated the \ac{s2r} matching approach (\Cref{sec:approach:reproduction}). For this RQ, we created another variant of our approach, \toolvarb, which did utilize the \acp{s2r} extracted by our approach, but used \recdroid's exploration algorithm instead of the Q-Learning approach defined in \Cref{sec:approach:reproduction}. Note that we used \recdroid's implementation for these variants since its implementation dependencies were more readily updated and its results in RQ2 showed it could be more easily adapted to run on newer and more modern subjects.  

To compute effectiveness, we ran the three approaches and two variants on each of the bug reports in the dataset. Following previous related works \cite{zhaoReCDroidAutomaticallyReproducing2019, fazziniAutomaticallyTranslatingBug2018}, each tool had a time limit of 3,600 seconds per bug report, and if the tool exceeded this time or threw an exception, then we considered the reproduction to have failed.  For each reproduction reported as completed by a tool, we manually replayed the identified event sequence to determine if it correctly reproduced the target failure.  Since our Q-Learning based reproduction is non-deterministic, we ran \tool and \toolvara three times on each bug report. If a tool succeeded at reproducing a given bug report at least once, then the reproduction of the said report was deemed a success.  To address RQ3, we measured the running time of all the approaches on each bug report. For \tool, we took the average of the three runs.

\subsubsection{RQ2 Effectiveness:}

Our results showed that \tool was effective in reproducing Android bug reports and outperformed the current state of the art tools (\ie \recdroid and \yakusu). Specifically, \tool was able to reproduce $57$ out of the total $77$ ($74\%$) bug reports, while \recdroid and \yakusu were only able to reproduce $37$ ($48\%$) and $9$ ($12\%$), respectively. Among the successfully reproduced ones by \tool, $98\%$ of the bug reports that were successfully reproduced were successfully reproduced more than once, which showcases the stability of the performance of our tool. Specifically, of the three runs of \tool, $1$ bug report was successfully reproduced once, $9$ twice, and $47$ all three times. Additionally, \tool also exhibited better performance on reproducing bug reports that were found to have missing steps. It was able to reproduce $32$ ($64\%$) out of a total of $50$ bug reports with missing steps. On the contrary, \recdroid was only able to reproduce $21$ ($42\%$) and \yakusu was only able to reproduce $3$ ($6\%$).  

One possibly confounding factor that we mention here is that \yakusu's implementation was tightly coupled with specific versions of Espresso and the Android OS.  In turn many of our more modern apps required more recent versions of Android.  While many of these apps could run with \yakusu, we confirmed with the authors of \yakusu that this issue could affect the generalizability of  \yakusu on new apps.



We further analyzed the reasons for the 20 bug reports that our tool was unable to reproduce.
Among them, eight reports could not be reproduced because the total number of missing steps in the bug report exceeded the limit of \ac{noop} actions set in \tool. For example, the reproduction of AnkiDroid-6432~\cite{ankidroid6432} required 26 steps, 20 of which were missed in the bug report. 
Two reports, Transistor-149~\cite{transistor149} and Materialistic-1067~\cite{materialistic1067}, could not be reproduced since the reproduction required pressing a button very quickly and the crash was non-deterministic. 
One report, FDroidClient-1821~\cite{fdroidclient1821} was failed to be reproduced because \tool failed to extract a \ac{s2r} from the bug report due to the failure of the underlying tool.
\tool failed to bridge the missed step, since the step is to click a specific button in a long list containing many other possible buttons. \tool failed to infer the correct button in this case.
The remaining nine bug reports could not be reproduced because the \ac{s2r} text description written in the bug report did not match with the reproducing \ac{ui} events. This happens because reporters may use their own words when describing the steps and the widgets do not have a meaningful description. In either case, even though the \ac{s2r} is given in the bug report, \tool could not find a match between the given \ac{s2r} and the correct \ac{ui} event.

We also analyzed the bug reports that our tool was not able to reproduce for all three runs.
Specifically, we found one bug report that our approach was only reproduced once and none of \recdroid and \yakusu could reproduce it.
We found that the reason is that the provided \acp{s2r} do not match with the \ac{ui} events.
Therefore, our approach spent a long time figuring out how to match these \acp{s2r} and was only managed to reproduce it once.
Among the eight bug reports that \tool successfully reproduced twice, the same reason was also found on four bug reports.
For the other four bug reports, \tool generates a \ac{ui} event sequence that triggers the same error message as the failure reported by the bug report but in a different way. We did not consider them as true successful reproductions. The potential reason is that \tool stops exploring the \ac{aut} once it found the \ac{aut} crashes with the specified error message. Therefore, it may trigger the crash when performing random exploration.

\subsubsection{RQ2.1\&2.2 Contribution of Different Components:}

Our results showed that \toolvara and \toolvarb both performed worse than \tool. \toolvara only reproduced $45$ bug reports and \toolvarb only reproduced $35$ bug reports. Note that \tool was able to reproduce all bug reports that \toolvara and \toolvarb were able to reproduce, with exception of one bug report that only \toolvara was able to reproduce. The difference in reproduction results between \tool and the two variants showed the contribution of the combination of both the \ac{nlp} techniques and the \ac{rl}-based exploration. In comparison to \recdroid, \toolvara reproduced eight more bug reports in total (i.e., $45$ to $37$), and \toolvarb reproduced two less in total (i.e., $35$ to $37$). Specifically, \toolvara succeeded on $14$ bug reports where \recdroid failed, and \recdroid succeeded on $6$ bug reports where \toolvara failed. Moreover, \toolvarb succeeded on $1$ bug report where \recdroid failed, and \recdroid succeeded on $3$ bug reports where \toolvarb failed.

We further investigated \toolvara's result to understand the reasons behind the changes in its performance compared with \tool. We observed that the reduced effectiveness of \toolvara can be primarily attributed to the inaccurate \acp{s2r} extracted by \recdroid. For example, when \recdroid misidentified an extra input \acp{s2r}, our \ac{rl}-based exploration method was not able to match it with any \ac{ui} event. However, even with mistakes in identifying \acp{s2r}, \toolvara still outperformed \recdroid by reproducing 8 more bug reports in total. This showed that, in many cases, our \ac{rl} matching algorithm was able to make up for the inaccuracy in \acp{s2r}.

We also explored the reasons why \toolvarb did not reproduce more bug reports and found the reasons to be twofold. First, \recdroid's exploration algorithm was found to be less effective in handling low-quality steps and missing steps. When a report contained poorly-written or missing steps, the algorithm took a long time to discover the correct matching, which often caused a timeout. Second, \recdroid's algorithm failed to leverage some useful information, such as \textit{resource id} or \textit{content description} for a \ac{ui} element from the \ac{ui} layout, during the exploration. The algorithm, instead, only utilized the displayed text on a given \ac{ui} element as a reference when matching it with an \ac{s2r}, which was not enough. In conjunction, these limitations prohibited \toolvarb from effectively utilizing the \acp{s2r} extracted by our approach.

\subsubsection{RQ3 Running Time:}

\tool was efficient in regards to running time in the reproduction of bug reports in comparison to \recdroid and \yakusu. On average, \tool spent $1,334$ seconds ($30$ on \ac{s2r} Extraction and $1304$ on \ac{s2r} Matching) on each reproduction, which was the fastest amongst all three techniques. In contrast, \recdroid and \yakusu spent $1,991$ ($4$ on  \ac{s2r} Extraction and $1,987$ on \ac{s2r} Matching) and $3,245$ seconds ($496$ on \ac{s2r} Extraction and $2,749$ on \ac{s2r} Matching), respectively, on each reproduction.

\subsection{Threats to Validity}

\subsubsection{External Validity}\label{sec:external_validity}
The primary threat to external validity is the representativeness of the bug reports in the evaluation dataset. 
We attempted to overcome this threat by using bug reports from related research works, which aided in avoiding biases that could potentially be introduced by our own search for reports.
Note that similar to \recdroid and \yakusu, our approach is unable to reproduce non-crash bug reports. 
Therefore, we did not collect non-crash bug reports in the evaluation dataset.



\subsubsection{Internal Validity}
One threat to the internal validity is the potential effect on the experimental results of the randomness of the \ac{rl} algorithm used by \tool for \ac{s2r} matching. To reduce this threat, we ran our tool and \toolvara three times and reported the running results for each run. An additional threat is potentially incorrect ground truth for RQ1. This threat was mitigated by having several authors individually define the RQ1 ground truth and come to a consensus on the finalized version. Note that neither \recdroid nor \yakusu carried out a similar evaluation, so this information was not provided in their online datasets. The last internal validity threat is regarding the issues we observed when running \yakusu on our dataset. These issues came from the limitations of \yakusu's implementation. As much as possible, we worked to set up compatible environments that could run the app and tool.

%% file: content/related_work.tex
\section{Related Work}

\recdroid~\cite{zhaoReCDroidAutomaticallyReproducing2019} and \yakusu~\cite{fazziniAutomaticallyTranslatingBug2018} are the most closely related works.
Their approaches rely on manually-crafted patterns to analyze bug reports and greedy-based exploration to discover the reproducing \ac{ui} events.
In contrast, our approach incorporates a combination of advanced \ac{nlp} techniques to analyze the bug report, which could extract \acp{s2r} more accurately. Moreover, our approach leverages \ac{rl} to guide the search for reproducing event sequences, which is more effective.  \Cref{sec:motivating_example} discusses both in more detail, and we compare against both \recdroid and \yakusu in the evaluation. 

There is some research work focusing on studying and analyzing Android bug reports.
Johnson~\etal~\cite{johnsonEmpiricalInvestigationReproduction2022} 
conducted an empirical study on 180 Android bug reports to identify challenges to reproducing Android bug reports.
Chaparro~\etal
~\cite{chaparroDetectingMissingInformation2017} 
conducted an empirical study on how users report observed behavior; \acp{s2r} and expected behavior; and identified discourse patterns used by reporters.
Based on these identified patterns, they designed an automatic tool to detect which information were missing in bug reports. However, none of them automatically extracted \acp{s2r} from the bug report or reproduced the report.
Chaparro~\etal developed \euler ~\cite{chaparroAssessingQualitySteps2019}, an automatic technique to assess the quality of \acp{s2r} in Android bug reports. \euler resolves \acp{s2r} from bug reports using simple grammar patterns. Different from \euler, our approach holistically analyzes the bug report's sentences to extract \acp{s2r}, which is more broadly applicable.

Several previous research works have focused on augmenting Android bug reports or facilitating the reporting process.
Liu~\etal proposed a machine learning based classifier, \maca
~\cite{liuAutomatedClassificationActions2020}, which classifies action words of \acp{s2r} into standard categories (\textit{click}, \textit{input} \etc) using both the textual information and the \ac{aut} information. 
\maca could potentially complement our technique by standardizing the action. However, \maca used simple grammar patterns to extract \acp{s2r}, which limits its capability when it cannot accurately parse the \acp{s2r}. \fusion, developed by Moran \etal~\cite{moranAutocompletingBugReports2015}, leveraged dynamic analysis to obtain \ac{ui} events of the \ac{aut} and help create more actionable events in bug reports during the testing stage. Fazzini \etal proposed \ebug~\cite{fazziniEnhancingMobileApp2022a} to assist reporters to write more accurate reproduction steps by using information from static and dynamic analysis of the \ac{aut} to predict the next step. Yang \etal proposed an interactive bug reporting system,  \burt~\cite{songInteractiveBugReporting2022}, which provides a guided reporting of essential bug report elements (i.e., the observed behavior, expected behavior, and
steps to reproduce the bug), instant feedback of problems with the elements, and graphical suggestions of the said elements. These three approaches mainly help improve the quality of the bug report at the moment when users write the report, but do not reproduce them. Our technique is complementary to these techniques by automatically reproducing the reports for developers.

Several works used \ac{rl} for automatically testing Android apps~\cite{vuongReinforcementLearningBased2018, adamoReinforcementLearningAndroid2018, panReinforcementLearningBased2020} or web applications~\cite{zhengAutomaticWebTesting2021}.
However, different from these work, we are the first work to adapt \ac{rl} to the Android bug report reproduction domain.

%% file: content/conclusion.tex
\section{Conclusion}
In this paper, we proposed a novel approach to automatically reproduce Android bug reports.
Our approach leverages advanced \acl{nlp} techniques to holistically and accurately analyze a given bug report and adopts \acl{rl} to effectively reproduce it.
The empirical evaluation shows that our approach is able to accurately extract reproduction steps from the bug report and effectively reproduce the bug report.
Our approach significantly outperformed state of the art techniques.